\documentclass[twocolumn]{article}

\usepackage{graphicx}
\usepackage{subfigure}
\usepackage{algorithm}
\usepackage{algorithmic}
\usepackage{hyperref}
\usepackage{url}
\usepackage{color}
\usepackage{mathtools}
\usepackage{amsfonts}

\usepackage{flushend}

\newcommand{\ornlthanks}{\thanks{This manuscript has been authored by a contractor of the U.S. Government under contract DE-AC05-00OR22725. Accordingly, the U.S. Government retains a nonexclusive, royalty-free license to publish or reproduce the published form of this contribution, or allow others to do so, for U.S. Government purposes.}}

\renewcommand{\P}{\ensuremath{\mathrm{P}}}

\begin{document}
%
\title{Nonparametric Bayesian Modeling for\\Automated Database Schema Matching\ornlthanks}
\author{Erik M.~Ferragut\\Oak Ridge National Laboratory
\and Jason Laska\\Oak Ridge National Laboratory}
\date{}
\maketitle
\begin{abstract}
\small\baselineskip=9pt
  The problem of merging databases arises in many government and
  commercial applications.  Schema matching, a common first step,
  identifies equivalent fields between databases.  We introduce a
  schema matching framework that builds nonparametric Bayesian models
  for each field and compares them by computing the probability that a
  single model could have generated both fields.  Our experiments show
  that our method is more accurate and faster than the existing
  instance-based matching algorithms in part because of the use of
  nonparametric Bayesian models.
\end{abstract}

\section{Background and Motivation.}
The trend health care, finance, and government sectors toward data
sharing has increased the need for data integration.  Furthermore,
organizations are mandated to integrate their data, whether due to a
corporate merger, legislated duties, international military efforts,
or disaster management.  A strong economic incentive exists for data
integration resulting from its benefits for anomaly detection, data
quality processing, fraud detection, and streamlining processing.

The data integration problem includes both schema matching and
coreference as subproblems.  Schema matching is the problem of
identifying {\em fields} \footnote{We consistently use the term field,
  but the terms attribute, column, and feature are also used in the
  literature.} that refer to the same concepts.  Coreferencing is the
problem of identifying {\em records} that refer to the same underlying
entity.  The difficulty of automatically attaining high quality
matches has motivated research to learn the schema matching using a
small number of coreferents~\cite{perkowitz1995category,
  chua2003instance, bilke2005schema}, to learn coreferents given a
matched schema~\cite{bellare2012active}, and to learn both schema and
coreferents simultaneously~\cite{wick2008unified}.  However, there is
a need for an out-of-the-box schema matching solution that is
independent of coreferencing.  This paper contributes to that goal.

Methods for automating schema matching have been explored in the
scientific literature~\cite{bernstein2011generic} and have been
included as part of business analytics tools (by, e.g., IBM, SAS,
Oracle, and Microsoft).  The majority of this work has focused on
using available metadata such as field names, on providing user
interfaces for manual field linking, and on developing effective
matchers as ensembles of individual matchers.  We review the related
work in Section~\ref{sec:related}.  Despite all of this previous work,
practical exercises of data integration continue to be largely manual
processes.

The primary contributions of this paper are
\begin{itemize}
\item a set of three nonparametric Bayesian model classes for use
  within a new probability-based schema matching framework
  (Section~\ref{sec:models}),
\item evidence that these model classes outperform
  existing instance-based matching scores (described in
  Section~\ref{sec:baselines}) in both accuracy and speed
  (Section~\ref{sec:results}), and
\item evidence that the improved performance is due at least in part
  to the use of nonparametric Bayesian models
  (Section~\ref{sec:anal}).
\end{itemize}

\subsection{Related Work.}
\label{sec:related}
A majority of previous work has been devoted to using metadata for
matching fields~\cite{bernstein2011generic}.  These methods include
exact and inexact~\cite{bilke2005schema} matching of field names,
synonym-based matching~\cite{embley2001multifaceted}, and other
language-based analyses~\cite{madhavan2001generic}.  These methods
assume a coherence between named fields and are likely to perform
poorly if the same data is called, say, Customer Name in one data set
and Guest ID in another.

Many existing machine learning methods attempt to learn how to match
names or other metadata using dictionaries or natural language
processing methods.  Those methods incur the additional burden of
obtaining a good good training set or solving the associated transfer
learning problem.  Either concern weakens the generalizability of any
proposed system, as the solutions to these problems may be domain
specific.

It should be noted that matchers are typically used in combination, as
this practice has been shown to be effective~\cite{do2002coma,
  bernstein2004industrial}.  However, to help clarify the impact of
our contributions, we focus on each instance-based matcher
separately.

Our approach is fully instance-based and ignores any metadata that
may be available.  Some previous work has been instance-based.
Instance-based methods use each field to produce a summary and then
compare the summaries.  These summaries tend to be the set or multiset
of values.  We compare our framework against the best instance-based
methods in the literature.

More recently, especially as researchers have shifted their focus from
databases to ontologies, additional emphasis has been placed on
exploiting the relationships among fields (also called concepts in the
ontology context), such as is-a and has-a relationships.  Because
these methods are applied to expert-developed ontologies (e.g.,
different anatomy ontologies) there are generally only a few available
instances for each field.  Methods exist to leverage known matched
instances for schema matching~\cite{drumm2007quickmig}.  Such matched
pairs provide a significant advantage in finding schema matches.  In
many applications, including typical cross-organizational data
integration efforts, the existence of common referents cannot be
assumed.  Furthermore, even if such common referents exist, finding
them is itself a highly challenging research problem.  Our method does
not depend on having coreferents.

\subsection{Baseline Methods.}
\label{sec:baselines}
Instance-based schema matching is generally pursued by defining
similarity or distance metrics between two fields, and then using
these scores to determine the matching decisions.  There are several
field matching scores that have been studied in the literature.  We
compare our method to five prominant and representative similarity
scores.  Two scores are based on set intersections and three scores
use the full multiset of counts.  The Jaccard Coefficient and the
Pointwise Mutual Information are described in
\cite{isaac2007empirical, wang2012instance}.  So-called corrected
versions are also described, but we will not discuss them here since
they consistently underperformed the uncorrected versions in all of
our experiments.  Kang and Naughton~(2003) introduce
information-theoretic measures based on mutual information and
entropy.  Jaiswal {\em et al.}~(2010) introduce the Euclidean
distance on the sorted normalized value counts.  Their use of the
distance on the sorted counts is meant to support detecting value
transformations, which we do not consider.  The natural alternative is
to use the Euclidean distance on the unsorted normalized value counts,
which we also include although they did not explicitly define or use
it.

To define the baseline methods, we use the following notation for a
fixed pair of match candidates.  Let $C$ and $D$ be sets of observed
values from the two fields, $N$ be the total number of observations,
including repititions, $p_i$ and $q_i$ be the proportion of
observations that were equal to the $i$-th distinct value, and $p'$
and $q'$ be $p$ and $q$, respectively, but each in decreasing order.
The names for the following statistics are chosen consistently with
the literature.

\begin{align*}
&{\text{Jaccard Coefficient}} & &\frac{|C \cap D|}{|C \cup D|} \\ 
&{\text{Pointwise MI}}  & &  \log_2 \frac{ |C \cap D| \times N }{|C| |D|} \\
&{\text{Entropy Difference}} & & \left| \sum_i p_i \log
  p_i - \sum_i q_i \log q_i \right| \\ 
&{\text{Unsorted Euclidean}} &  & \sum_i (p_i - q_i)^2 \\
&{\text{Sorted Euclidean}} &  & \sum_i (p'_i - q'_i)^2 \\
\end{align*}
The first two scores give larger values for more likely matches,
whereas the other scores (the last three) give smaller values for more
likely matches.  Two additional similarity scores, Jensen-Shannon and
log likelihood, are considered in the literature, but we do not
include them here since they require the set of observed values to be
the same for both fields, which is almost never the case.

The references apply the metrics within more complex matching schemes
using ensemble scores~\cite{peukert2012self}, limits on the number of
matches per field~\cite{melnik2002similarity}, and collective
optimization~\cite{papadimitriou2012taci}.  For clarity, we focus on
the more straightforward, though harder, problem of deciding whether
two sets of instances should be matched or not, without regard to the
other available information and restrictions.

The variety of set-based and multiset-based similarity functions
studied have two main shortcomings.  First, they are very coarse in
the sense that a lot of information regarding similarities between
values is discarded.  Second, they tend to be computationally very
expensive.  In many cases, these methods require comparing every
value in one field to every value in the other, which is work on the
order of the number of distinct values for each pair.

Non-multiset-based methods have been explored in the literature. For
example, Jaiswal {\em et al.}~model continuous variables by Gaussian
mixtures.  Other research has pursued value
classification~\cite{lambrix2008literature} and clustering
approaches~\cite{li2000semint, algergawy2011clustering}.  We do not
compare directly to these alternative methods, primarily because they
are computationally prohibitive for large data sets.

\section{Methods and Technical Solutions.}
\label{sec:framework} \label{sec:models}


We explore the hypothesis that {\em using probabilistic models that
  meet certain simplicity constraints enable both greater accuracy and
  greater computational efficiency}.
We view field values as being generated according to
probabilistic models, which allows for explicit computation of the
probability of a match given the observed data.  

The probabilistic field matching framework uses a collection of model
classes to (1) train models based on string instances observed in each
field, and (2) compare models by computing the relative likelihood
that both fields were generated from the same models. 


The process of matching fields is then as follows.  First, initial
models are created for each field for each model class and then
updated efficiently with the data from that field by computing the
sufficient statistics that determine the parameters (see next section
for model-specific details).  Second, the probability of a match for
each pair is computed.

We pose the field match problem as a probability computation.  For any
pair of fields, we assume two mutually exclusive and exhaustive
possibilities, either (1) there was one model generating both fields,
or (2) the fields were generated by independent models.  We denote the
former as $S$ and the latter as $\neg S$.  By Bayes' rule, the
probability of a match for field data $X$ and $Y$ is then $\P( S \mid
X + Y ) =$
\begin{equation} \label{eq:probmatch}
\frac{\P( X+Y \mid S) \P(S)}{
\P( X+Y \mid S) \P(S) + \P( X +Y \mid \neg S) \P(\neg S )},
\end{equation}
with $\P(X+Y \mid \neg S) = \P(X) \P(Y)$, since it uses independent
models for $X$ and $Y$.  This computation is done for each model class
separately.  Generally, the match scores will form the basis for
follow-on processing for data integration or other related purposes.

We created and implemented three probabilistic model classes for
string generation satisfying the design constraints.  Each model class
uses the Chinese Restaurant Process (CRP) in modeling the collection of
all possible strings.  The CRP may be viewed as a principled
generalization of a Dirichlet distribution to infinite dimensions that
maintains exchangeability and has no zero probabilities (provided that
the base distribution has none).  The use of the CRP, while not new in
the literature, is somewhat unusual due to the fact that we are
working with a probability mass function rather than a probability
density function.  



The possible values for a field will only rarely be known in advance.
Consequently, typical methods for modeling categorical data, such as
multinomial distributions, will usually not apply.  Instead, Dirichlet
processes are appropriate.  A Dirichlet process is a stochastic
process that generates a Dirichlet distribution.  Loosely speaking,
the Dirichlet process provides a way to select a finite number of
categories and to build a multinomial for it, while still allowing for
the possibility of new events.  Although a full technical description
of Dirichlet processes is beyond the scope of this paper, only an
understanding of the probability scoring described in~\eqref{eq:crp}
will be needed.  A thorough survey is available
from~\cite{teh2010dirichlet}.

One useful metaphor for the Dirichlet process is the Chinese
Restaurant Process (CRP).  The CRP models a series of arrivals at a
restaurant that has a countably infinite number of tables with
unlimited seating.  A new customer either chooses a table with
probability proportional to the number of customers already seated at
the table or else picks a new table with probability proportional to
the concentration parameter $\alpha$.  The first customer necessarily
picks a new table.  Every time a new table is selected, a label is
generated according to a specified base distribution.  In the typical
treatment, the base distribution is assumed to be non-atomic, so that
the probability of a subsequent new table generating a repeated label
is zero.  In the Atomic CRP, this assumption is waived.  The
fundamental theory remains the same.  Without loss of generality, a
customer at a new table that chooses a repeated label is moved to the
table already having that label.

The probability of getting $m_i$ instances of value $x_i$ where the $x_i$
are generated according to a base distribution $H$ in a CRP with parameter
$\alpha$ is given by
\begin{equation} \label{eq:crp}
    \frac{ \Gamma(\alpha) }{   \Gamma( \alpha + \sum_i m_i ) }
 \prod_i \frac{\Gamma( \alpha H(x_i) + m_i)}{
 \Gamma( \alpha H(x_i) )},
\end{equation}
where $\Gamma$ is the standard Gamma function.  By comparison, the
non-atomic CRP is obtainable from~\eqref{eq:crp} by replacing $H(x_i)$
with 1.  Of course, the two versions of the CRP model slightly
different data since only the ACRP accounts for the
labels.  

\subsection{Discrete Model.} \label{subsec:disc}

The discrete model class is the simplest of the three model classes.
The table labels from the Atomic Chinese Restaurant Process are the
set of distinct values in the field.  As a base distribution ($H$ in
\eqref{eq:crp}), we choose the following string generation process.
First, select a string length according to a Poisson distribution with
fixed $\lambda$.  Next, generate that many uniformly chosen characters
from the alphabet.  A discrete model must track the entire multiset of
observations.

To compute the probability of the data, we use the standard Bayesian
paradigm. For $M$ the model class and strings $x_1, x_2,
\dots, x_n$,
\begin{align*}
    \P( x_1, &x_2, \dots, x_n \mid M ) = \\
&\prod_{i=1}^n \P( x_i \mid M, x_1, x_2, \dots, x_{i-1} ) 
\end{align*}
where each term uses the parameters computed from the previous data to
calculate the probability of the next observation.  The joint
probability is exchangeable; it does not matter in what order the data
are observed.  Moreover, the joint probability can be computed
directly and quickly from the parameters in terms of the Gamma
function, as per~\eqref{eq:crp}.

\subsection{Positional Model.} \label{subsec:pos} 

We define a positional probability model that generates strings in two
steps.  First, a length $\ell$ is sampled from an Atomic Chinese
Restaurant Process with a Poisson base distribution.  Second, $\ell$ characters
are sampled from the first $\ell$ character distributions, which are
modeled with separate uniform Dirichlet distributions on a fixed known
alphabet $A$.  The model parameters for this model are computed by
counting the number of strings of each length and the number of times
each character is observed at each position.  Let $n_\ell$ be the
number of strings of length $\ell$.  For $j = 1, 2, \dots$ and for $a
\in A$ a character in the fixed alphabet, let $c_{j,a}$ be the number
of times character $a$ was observed in position $j$.  The probability
of the $i$-th string $x_{i,1} x_{i,2} \dots x_{i,\ell_i}$ for $\ell_i$
its length is the probability of the length times the probabilities of
the characters.

\begin{align} \label{eq:thisformula}
   \P( x_{i,1} &x_{i,2} \dots x_{i,\ell_i} \mid n, c, \alpha, \lambda, \beta) = \\
&\frac{ n_{\ell_i} + \alpha \text{Pois}_\lambda( \ell_i ) }{ n + \alpha }  
\prod_{j=1}^{\ell_i}
 \frac{ c_{j, x_j} + \beta }{  n_{\geq j} + |A| \beta }
\end{align}
where $\alpha$ is the CRP strength parameter for the length, $\lambda$
is the parameter for the CRP's Poisson base distribution, $\beta$ is
the Dirichlet prior for all character multinomials, $|A|$ the fixed
alphabet size, and $n_{\geq j}$ is for the number of strings observed
with length at least $j$.  Also, an empty product is 1 by convention.
Equation~\eqref{eq:thisformula} is useful for computing the
probability of a single string and could be used to compute the joint
probability of the data, but a simpler product using the Gamma
function is possible.
\begin{align} \label{eq:posprob} \nonumber
       \P( x_1, x_2, &\dots, x_n \mid M ) =
       \frac{ \Gamma(n_\ell + \alpha \text{Pois}_\lambda(\ell)) \Gamma(\alpha)  }{ 
         \Gamma(\alpha \text{Pois}_\lambda(\ell)) \Gamma( n +
         \alpha) }  \\
\prod_{j=1}^{\max\ell} &\prod_{a \in A}
 \frac{ \Gamma(c_{j, a} + \beta) }{\Gamma(\beta)} 
 \frac{ \Gamma( |A| \beta ) }{ \Gamma(n_{\geq j} + |A| \beta) }
\end{align}
where $M$ denotes the model class.

The positional model must track the number of times each character was
observed in each position as well as the number of strings of each
length.  The parameters for the combined data from two columns are the
sum of the parameters learned from each column.  The joint probability
is computable in time proportional to the number of parameters, rather
than in time proportional to the amount of data, a fact that is
especially important when computing the probability that two fields
were generated from the same model.  That is, $c^{(X+Y)}_{j,a} = c^{(X)}_{j,a} +
c^{(Y)}_{j,a}$ for all $j$ and $a$.

\subsection{Apositional Model.} 
\label{subsec:apos} 
The apositional model is a simplification of the positional model.  In
generating strings, it chooses the length in the same way.  However,
the characters are produced with probabilities that are independent of
the position.  The probability of data given the model can be computed
using the same formulas, but in place of $c_{i,a}$ we use
\[
   c'_a = \sum_i c_{i,a}.
\]
That is, the counts are combined across positions.  Alternatively, it
can be viewed as imposing a parameter equality across the positions.
To combine two models' parameters $c$ and $n$, we again simply add
them.

\section{Empirical Evaluation.}
\label{sec:experiments}
We tested our nonparametric Bayesian schema matching approach with a
number of experiments. The experimental procedures followed are
described in Subsection~\ref{subsec:ed}.  The data sets are described
in Subsection~\ref{subsec:data}.

\subsection{Experimental Design.}
\label{subsec:ed}
To measure the accuracy of the schema matching, we performed a
subsample self-match.  First, the data sets are cleaned.  Strings are
normalized to a 64 character alphabet by capitalizing all letters and
replacing any character that is not a digit or a punctuation mark (in
particular, period, comma, colon, semi-colon, slash, backslash, quote,
apostrophe, back tick, bracket, parenthesis, plus, minus, exclamation
mark, question mark, dollar sign, percent, ampersand, asterisk,
underscore) with a placeholder symbol.  Fields in which 99\% of the
values were the same were removed; this included removing empty fields
since all their values were empty strings.  Second, we create two
subsample data sets with the first and last $n$ records.  Third, we
take ground truth to be that the $i$-th field of one sample only
matches the $i$-th field of the other sample.  This is not a perfect
assumption in the sense that multiple fields can represent the same
type of data (e.g., dates).  However, since subsample self-matching
enables a consistent comparison between methods and across data sets
with reliable ground truth, subsample matching has become
standard~\cite{jaiswal2010uninterpreted}.  Fourth, we run the baseline
schema matching methods and our probabilistic field modeling approach
to match fields between the two subsamples.  Every pair of one field
from the first subsample and one field from the second is then an
example that is scored and compared to ground truth.  Given $d$
fields, we obtain $d^2$ examples, of which $d$ are matches and $d^2 -
d$ are non-matches.  We compute standard machine learning measures of
success, including the ROC curve and the area under the ROC curve.


In the literature it is common to downselect the number of features
used in pairwise match computations to ten to thirty. Furthermore,
they often include constraints on the number of matches allowed for
any given field, such as limiting every field in one data set to at
most one match in the other data set~\cite{dhamankar2004imap}.  These
sorts of assumptions, while often reasonable, will not be valid in
general.  Consequently, we preferred to test our method in the more
realistic and more difficult case where no constraints are known.

All models were chosen using the same priors and base distributions.
The parameter $\alpha$ for the Chinese Restaurant Process for the
string length was 3.0.  The mean string length $\lambda$ was 4.0.  The
prior $\beta$ for the character distributions was 3.0.

\subsection{Data.}
\label{subsec:data}
We tested our schema matching algorithm on four different data sets:
Census, Loans, Mix Market, and NPPES.  Table~\ref{tab:datasizes}
provides the number of records, the original number of fields, and the
number remaining after removing empty and nearly constant fields.

\begin{table}[tb]
\caption{Data Set Sizes\label{tab:datasizes}}
\vskip 0.15in
\begin{center}
\begin{small}
\begin{sc}
\begin{tabular}{lrrr}
\hline
 & & \multicolumn{1}{c}{Orig.} & \multicolumn{1}{c}{Filtered} \\
Data Set & \multicolumn{1}{c}{Records} & 
\multicolumn{1}{c}{Fields} & \multicolumn{1}{c}{Fields} 
\\ \hline
Census & 581746 & 118 & 101 \\
Loans & 147638 & 101 & 100 \\
Mix Market & 14736 & 83 & 81 \\   
NPPES & 1308299 & 329 & 101 \\
\hline
\end{tabular}
\end{sc}
\end{small}
\end{center}
\vskip -0.1in
\end{table}

The Census data comes from the 1990 Census, and is provided by The
United States Census Bureau provides the 1990 Public Use Micro Sample
5\% data from California\footnote{We used the California file
  available from
  \url{http://www2.census.gov/census_1990/1990_PUMS_A/}.}.  The fields
are fixed-width numeric encodings according to a data dictionary.  We
did not replace the codes with their dictionary values.

The Loans data contains complete loan information, such as loan status
and payment information\footnote{The Loans data is available from
  \url{https://www.lendingclub.com/info/download-data.action} by
  clicking on the ``2012--present load data'' button.}.  The data come
from Lending Club, an online financial community that matches
individual borrowers with individual lenders and has loaned over \$2.5
billion dollars to date.  The fields include durations, amounts,
percentages, and dates.

The Mix Market data come from Microfinance Information Exchange, Inc.,
a non-profit organization that manages and provides information about
financial institutions engaged in microlending, the practice of making
high-impact small-capital loans to disadvantaged borrowers.  We used
their freely available Basic MIX MFI Data Set\footnote{To download the
  data, click on the ``Download Basic MIX MFI Data Set'' button at
  \url{http://www.mixmarket.org/profiles-reports} and follow the
  directions.}.  In contrast to the Census data, the Mix Market data
presents a wide variety of field types.  For example, it includes and
ID number, an organization name, currency type and amounts
(separately), years, dates, and percentages.

The NPPES data is a large data set managed by Center for Medicare and
Medicaid Studies containing public information about registered
Medicare service providers\footnote{The most recent monthly full
  replacement file is available from
  \url{http://nppes.viva-it.com/NPI_Files.html}.}.  NPPES has by far
the largest number of records of our three data sets, which allows us
to test our algorithms on especially large data sets.  The data
contained are diverse, although most of it is comprised of contact
information, such as name, phone number, and address fields.  Of
special interest within the health care domain is the problem of
handling the wide diversity of provider IDs that appear in various
fields, especially the 73 different ``Other Provider Information''
fields.  These fields are used to collectively capture a list of
values, which challenges the accuracy of any automated schema matching
solution.

\subsection{Results.}
\label{sec:results}
\begin{figure}[tb]
\begin{center}
\includegraphics[width=3in]{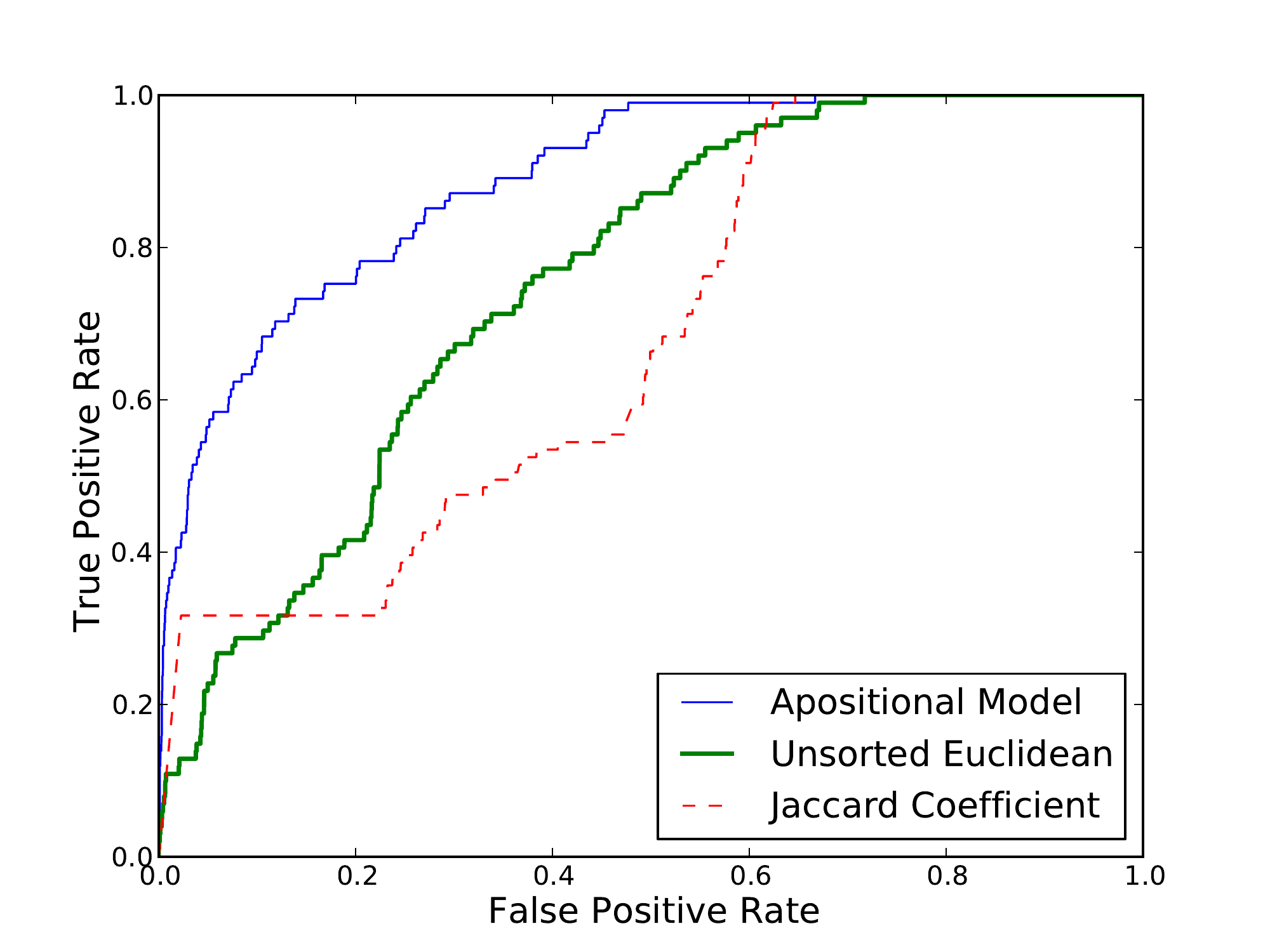}
\end{center}
\caption{The ROC curve for the apositional field model (the best
  probabilistic model in this experiment) is superior to
  both the set-based and multiset-based schema matching approaches for
  most false positive rates.  This comparison is for subsamples of
  size 50,000 within the NPPES data set.\label{fig:npibest50000}}
\end{figure}

Fig.~\ref{fig:npibest50000} shows that the best\footnote{Although
classifiers should properly be compared on numerous factors, we use
the term ``best'' as a convenient way to refer to the model with the
largest area under the ROC curve (AUC).} classifier on the
NPPES data set with samples of size 50,000 is the apositional
probability model.  The best multiset-based model is the unsorted
Euclidean score.  The set-based scores performed the worst, but the
better one is the Jaccard coefficient.  This pattern, in which the
probabilistic models perform the best and the set-based methods the
worst was consistent across all experiments.

In many cases the apositional model was the best of the probabilistic
models, followed by the positional and discrete models.  One exception
was in the Census data with samples of size 500 where the discrete
model was the best, and the apositional was the worst.
Table~\ref{tab:auc} summarizes the AUC statistics for the
probabilistic models across the four data sets with samples of size
5000.  {\em In every case, the best model is a nonparametric Bayesian
  models.}


The Table also shows a clear performance gap between set-based and
multiset-based methods.  The Sorted Euclidean and Unsorted Euclidean
scores are consistently better than Entropy Difference or the set-based
methods. This observation replicates the findings
in~\cite{jaiswal2010uninterpreted}.

\begin{table}[tb]
  \caption{AUC for Subsamples of Size 5000.\label{tab:auc}}
\vskip 0.15in
\begin{center}
\begin{small}
\begin{sc}
\begin{tabular}{lrrrr}
\hline
 & \multicolumn{4}{c}{Data Set} \\
Model & \multicolumn{1}{c}{Census} & \multicolumn{1}{c}{Loans} & 
\multicolumn{1}{c}{Mix} & \multicolumn{1}{c}{NPPES}  
\\ \hline
Apositional   & 0.88 & 0.89 & 1.00 & 0.88  \\
Positional    & 0.87 & 0.85 & 0.99 & 0.87 \\
Discrete      & 0.91 & 0.87 & 0.87 & 0.79
\\ \hline 
Sorted Eucl.  & 0.86 & 0.68 & 0.98 & 0.74  \\
Unsorted Eucl.& 0.86 & 0.71 & 0.98 & 0.74  \\
Entropy Diff. & 0.82 & 0.69 & 0.93 & 0.70
\\ \hline 
Jaccard Coef. & 0.76 & 0.57 & 0.65 & 0.67  \\
PMI & 0.67 & 0.61 & 0.60  & 0.60  \\
\hline
\end{tabular}
\end{sc}
\end{small}
\end{center}
\vskip -0.1in
\end{table}


The apositional and positional probabilistic models, in addition to
often being the best performing, were also significantly faster.  One
reasonable way to judge the speed of each method is to count the
number of parameters it uses.  The model training and the field
comparison both require work on the order of the number of parameters.
This is true also for set-based and multiset-based methods if we take
their parameters to be the set and the multiset, resp.  Both of these
as well as the discrete model have as many parameters as there are
distinct values in the field.  Table~\ref{tab:speed} lists the
average number of parameters across all fields for NPPES and
for each model.  It shows that the apositional and
positional models computationally scale far better.

\begin{table}[tb]
\caption{\label{tab:speed}Average Number of Model Parameters}
\vskip 0.15in
\begin{center} \begin{small} \begin{sc}
\begin{tabular}{lrrr}
\hline
 & \multicolumn{3}{c}{Subsample Size}\\
Model        & 500 & 5000 & 50000
\\ \hline
Apositional   & 45  & 62   & 77     \\
Positional    & 214 & 411  & 667    \\ 
All Others     & 169 & 1338 & 10641  \\
\hline
\end{tabular} 
\end{sc}
\end{small}
\end{center}
\vskip -0.1in
\end{table}

\section{Significance and Impact.}
\label{sec:anal}
In this section, we show that using more data helps, but only
marginally, especially in comparison to the difference in performance
between methods.  We then show that the training for the positional
and apositional models allow for inference of character-level value
patterns.  Finally, we show by experiments that the success of our
approach is attributable (at least in part) to the properties of
nonparametric Bayesian models.

\subsection{Sensitivity to Data Size.}

We conducted experiments to examine the sensitivity of the determined
AUC to changing data size. Table~\ref{tab:size} shows that the AUC for
the apositional and positional models does not appreciably change with
an increase in data size.  Also, the difference in performance between
models is significantly larger than the gains in performance from a
100-fold increase in data size.  The performance difference in the
apositional and positional models is especially surprising when
considering that they use far fewer paramters (see
Table~\ref{tab:speed}).

\begin{table}[tb]
\caption{\label{tab:size}AUC for NPPES with Varying Subsample Sizes.}
\vskip 0.15in
\begin{center} \begin{small} \begin{sc}
\begin{tabular}{lrrr}
\hline
 & \multicolumn{3}{c}{Subsample Size} \\
Model         & 500  & 5000 & 50000
\\ \hline
Apositional   & 0.89 & 0.88 & 0.89 \\
Positional    & 0.87 & 0.87 & 0.87 \\
Discrete      & 0.78 & 0.79 & 0.83
\\ \hline
Sorted Eucl.  & 0.74 & 0.74 & 0.75 \\
Unsorted Eucl.& 0.73 & 0.74 & 0.75
\\ \hline
Entropy Diff. & 0.69 & 0.70 & 0.72 \\
Jaccard Coef. & 0.64 & 0.67 & 0.68 \\
PMI           & 0.59 & 0.60 & 0.61 \\
\hline
\end{tabular} 
\end{sc}
\end{small}
\end{center}
\vskip -0.1in
\end{table}



\subsection{Pattern Inference.}

The positional and apositional models extrapolate based on
character-level similarities between values, such as learning formats
and other patterns, without having to make a new model for each
pattern. In contrast, set and multiset methods cannot.  For example,
the positional model has learned the structure of date fields in the
Loans data set. In particular, the seventh position is a hyphen for
correctly coded values.  Similarly, the positional model also learned
that the first character of the NPI Code field is always the digit 1.
Additionally, the apositional model learns that the majority of
characters in a ZIP Code field are digits. The fact that it has
non-digits suggests a data entry error.  Using positional and
apositional models allows for the construction of system data quality
checks.

\subsection{Effect of Bayesian Computations and CRP.}
The approach outlined in this paper is focused on a probabilistic
framework and the models used within that framework.  We have
stressed that the models follow the Bayesian paradigm in which the
probability of the data is computed, at least theoretically, one
observed value at a time.  This approach motivated the use of the
Chinese Restaurant Process since it follows the paradigm while also
allowing for arbitrarily many different values, even if they are not
known in advance.

One commonly used approach that contrasts with the Bayesian paradigm
is the Maximum Likelihood Estimation (MLE) paradigm, wherein the data
are scored according the model that maximizes their likelihood.  In
using the probabilistic framework, we could have adopted a parametric
and non-Bayesian approach where the parameters for the models are
learned from the data.  In this subsection, we consider three MLE
versions of the three probabilistic models we used.  These approaches
compute the same parameters, but compute the probability of the data
by having the probability of an event $x$ as the proportion of
previous observations that were $x$.  For the apositional and
positional models, we also compute the average length of the strings
and use that as the mean for the Poisson that generates the string
lengths.  The character probabilities are set according to the
proportion of observations.

\begin{figure}[tb]
\begin{center}
\includegraphics[width=3in]{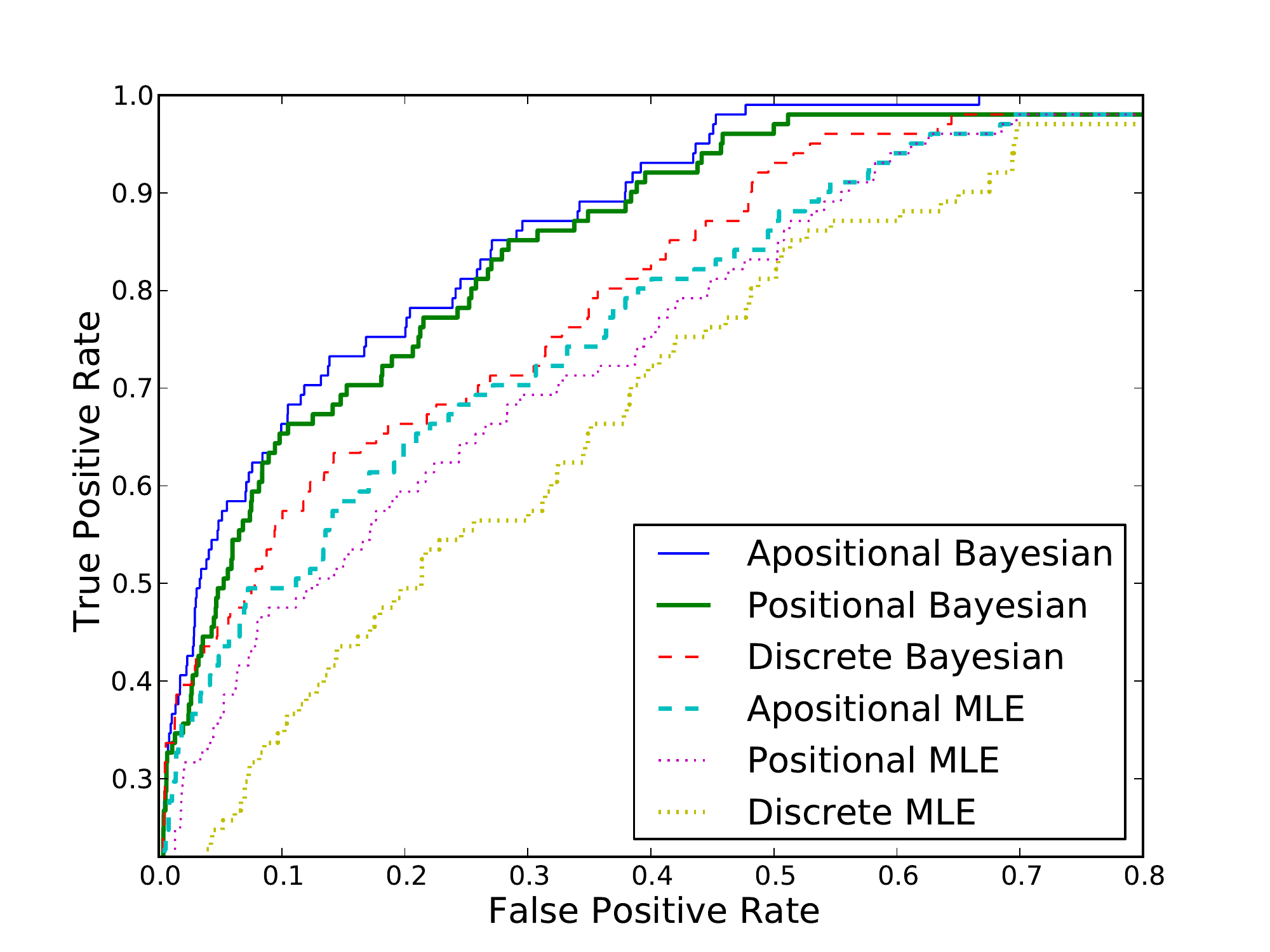}
\end{center}
\caption{The ROC curve for the nonparametric Bayesian models show
  better field matching than the Maximum Likelihood Estimation
  approaches for subsamples of size 50,000 within the NPPES data
  set.\label{fig:mleandpm}}
\end{figure}

By comparing the MLE versions and the Bayesian versions of the
probabilistic models, we were able to show that the Bayesian versions
generally perform better.  Fig.~\ref{fig:mleandpm} compares these
models on the NPPES data set with subsamples of size 50,000; these
curves represent the most competitive MLE results obtained.  We
conclude that the nonparametric Bayesian versions of the probabilistic
models attain better performance than the MLE versions.

We explain this difference by considering (1) the MLE training
process, (2) Bayes' rule with the MLE models, and (3) the usefulness
of the Chinese Restaurant Process.  The MLE training process uses the
data both to build the model and compute its probability.  Since we
are using the probability as an important component in our
classification, this approach could lead to overtraining.  Also, the
probability of two fields coming from the same model (i.e., $\P(S)$)
will always be lower than the probability of two fields coming from
different models in the MLE paradigm.  This is not true for the
Bayesian paradigm, and suggests that the MLE paradigm is not correctly
addressing the similarity question.  Finally, the Chinese Restaurant
Process was not part of the MLE models.  Consequently, the probability
of data with, say, a fixed length will be greatly underestimated.
This can result in undercounting information from length distributions
when assessing the field match quality.  Collectively, these
differences help to explain the better performance of nonparametric
Bayesian models.

\section{Conclusion and Future Work.}
This paper has introduced probabilistic field modeling, a novel
framework for schema matching that builds probabilistic models for
each field and uses the models to make determinations about which
fields should be matched.  We showed that this approach leads to more
accurate schema matching than existing instance-based methods.
Moreover, except for the discrete model class, it is computationally
faster due to not needing to retain the full set or multiset of
values.  We then showed that model training for positional and
apositional models allows the the system to learn patterns that are
typical for field values.  Finally, we showed that the imporved
performance is due in part to the use of nonparametric Bayesian
models.



This paper has shown that probabilistic field modeling can make a
significant contribution to the overall schema integration problem,
which will support business and government efforts to streamline their
data operations.  In addition to the commercial and government
applications, there are a number of significant scientific impacts of
using probabilistic field modeling for schema matching.  Development
of a probabilistic understanding of structured heterogeneous data may
have applications outside of schema matching.  For example, it should
be useful in characterizing typical and atypical data, in identifying
data quality issues, in discovering anomalies within a data set, and
in synthesizing realistic privacy-preserving proxy data.  In
conclusion, we have shown that our nonparametric Bayesian field
modeling framework has the potential to become an essential tool for
future heterogeneous data applications.

\bibliographystyle{plain}
\bibliography{../schema-match-lit/full-schema-matching}

\end{document}